# ARTICLE

## Rheological properties and shear-induced structures of ferroelectric nematic liquid crystals

Ashish Chandra Das,[a,b] Sathyanarayana Paladugu [b] and Oleg D. Lavrentovich *[a,b,c]

Recently discovered ferroelectric nematic ($N_F$) liquid crystals are fluids with a polar orientational order. The electric polarization vector can be aligned by an electric field and by surface anchoring. Here, we explore how the polarization field and effective viscosity of the $N_F$ materials are affected by shear flows. We explore three $N_F$ materials, abbreviated RM734, DIO, and a room-temperature FNLC919, all of which exhibit a paraelectric nematic (N) and the $N_F$ phase. All materials show an increase of the effective viscosity upon cooling, with an Arrhenius behavior in broad temperature ranges except near the phase transitions. In DIO and FNLC919, the antiferroelectric $SmZ_A$ phase separating the N and $N_F$ phases shows a strong dependence of the effective viscosity on the shear rate: this viscosity is lower than the viscosity of the N and $N_F$ phases at high shear rates ($\dot{\gamma} = 500\ s^{-1}$) but is much higher when the shear rate is low, $\dot{\gamma} = 2.5\ s^{-1}$. The behavior is associated with the layered structure of the $SmZ_A$ phase. All mesophases in all three materials exhibit shear-thinning behavior at low shear rates ($< 100\ s^{-1}$) and a nearly Newtonian behavior at higher shear rates. In terms of alignment, we observe three regimes in the N and $N_F$ phases: flow-alignment at low shear rates, $\dot{\gamma} < 10^2\ s^{-1}$, log-rolling regime with the director and polarization along the vorticity axis at $\dot{\gamma} > 10^3\ s^{-1}$, and polydomain structures at intermediate rates. In the flow-aligning regime, the $N_F$ polarization does not tilt away from the shear direction, which is in sharp contrast to the flow-induced tilt of the N director. The effect is attributed to the avoidance of splay deformations and associated space charge in the flowing $N_F$. The temperature and shear rate dependencies of the viscosity and the uncovered shear-induced structural effects of $N_F$ advance our understanding of these materials and potentially facilitate their applications.

## 1. Introduction

A nematic (N) liquid crystal (LC) formed by achiral rod-like molecules shows a uniaxial long-range orientational order but a lack of positional order. The molecules align along a common direction described by a unit vector $\hat{\mathbf{n}}$, called the director, with the property $\hat{\mathbf{n}} \equiv -\hat{\mathbf{n}}$, which makes the material paraelectric. Recent synthesis and characterization of liquid crystal established the existence of the ferroelectric nematic ($N_F$), in which rod-like molecules with large longitudinal electric dipoles align in a polar fashion along the director, resulting in producing a spontaneous macroscopic electric polarization **P**.[1-4] In the $N_F$ phase, the polarization is strong, $P \sim 6 \times 10^{-2}$ C/m², enabling electro-optic response to electric fields as small as $\sim 10^2$ V/m, a thousand times smaller than those used to reorient nonpolar N.[4]

Orientation of a paraelectric N is strongly affected by flows. Over the last few decades, significant progress has been achieved in understanding the shear-induced structures and rheology of the paraelectric N.[5-9] These materials exhibit different modes of response to the applied shear, such as flow-alignment,[5,6] tumbling,[10] log-rolling,[11,12] and kayaking.[7,8] For example, well-studied nematics MBBA[13] and 5CB[14] show a flow alignment, i.e, $\hat{\mathbf{n}}$ aligns parallel to the shear plane formed by the velocity and its gradient. In contrast, 8CB, formed by molecules with a slightly longer aliphatic tail as compared to 5CB and exhibiting a smectic A (SmA) phase in addition to the N, shows a tumbling behavior, with $\hat{\mathbf{n}}$ rotating in the shear plane and realigning along the vorticity direction perpendicular to the shear plane.[15]

In contrast to the paraelectric N, very little is known about the rheological behavior of the $N_F$. One should expect a rich plethora of flow phenomena, thanks to the presence of spontaneous electric polarization, its spatial variations, intrinsic to the fluid nature of the $N_F$, and strong coupling to the electric fields.[4,16] For example, Dhara et al. reported an increase in the effective viscosity of the $N_F$ material, abbreviated RM734, in the presence of an electric field.[17] Even stronger electroviscous response was reported recently by Nishikawa et al. for another $N_F$ material, abbreviated DIO.[18] However, the structural response of $N_F$ to shear, its flow regimes, dependence of the effective viscosities on shear rate and temperature, proximity of phase transitions, etc., remain underexplored.

In this work, we perform comparative rheological studies of the N, $N_F$ phases, as well as the intermediate phase separating the N and $N_F$, in three different materials, RM734,[1] DIO,[2] and FNLC919.[19] Namely, we measure the effective viscosity as a function of temperature and shear rate and determine the structural response to shear flows using a plate-plate rheometer equipped with an in situ polarizing optical microscope (POM). Effective viscosity increases as the temperature is lowered and the material transitions from the N to the $N_F$ phase. Far away from the phase transition temperatures, the viscosity follows the Arrhenius behavior but increases sharply near the transition points. As a function of the shear rate, both the N and $N_F$ phases show a strong shear-thinning at shear rates below 100 $s^{-1}$, and a nearly Newtonian behavior at higher shear rates. The structural response of the N and $N_F$ phases in all materials show three regimes as a function

[a.] *Material Science Graduate Program, Kent State University, Kent, OH 44242, USA.*
[b.] *Advanced Materials and Liquid Crystal Institute, Kent State University, Kent, OH 44242, USA*
[c.] *Department of Physics, Kent State University, Kent, OH 44242, USA*
*Email: olavrent@kent.edu*

of shear rate: flow alignment at low rates, $\dot{\gamma} < 10^2$ s$^{-1}$, log-rolling at $\dot{\gamma} > 10^3$ s$^{-1}$, and polydomain structures at intermediate rates. In the flow-aligning regime, the N director tilts away from the shear direction, while in the $N_F$ phase, such a tilt is absent. This important difference is caused by the avoidance of splay deformations and associated space charge in the $N_F$ phase.

## 2. Experimental methods

### 2.1 Materials

We explore three different $N_F$ materials, abbreviated RM734,[1] DIO,[2] and a room-temperature FNLC919.[19, 20] The phase sequences for these materials are presented in Fig. 1. RM734 was purchased from Instec, Inc. (purity better than 99 wt%). DIO was synthesized as described previously by Nishikawa et al.[2] A room-temperature $N_F$ material FNLC919 is provided by Merck KGaA (Darmstadt, Germany).

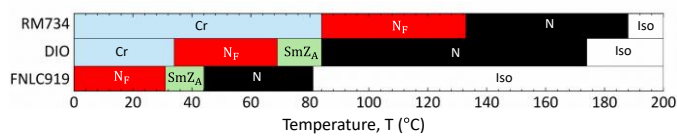

**Fig. 1**: Phase sequences of RM734, DIO, and FNLC919

The N and $N_F$ phases in all three materials are separated by an intermediate phase. Its nature is still debated, prompting us to present a brief overview. It was first observed in DIO,[2, 21-23] then in RM734,[24] and in other materials.[23, 25-28] This phase was labeled as $M_2$,[2] $N_x$,[22] $SmZ_A$,[29] $N_S$,[30] and $M_{AF}$;[31] the abbreviations reflect the perceived structure of this intermediate phase. The first report by Nishikawa et al. on DIO[2] pointed out that the transition upon cooling from the N to the new phase known nowadays as the $N_F$, happens in two steps. Brown et al.[21] and Erkoreka et al. suggested that the phase is antiferroelectric.[22] Subsequent synchrotron-based small and wide-angle X-ray scattering (SAXS/WAXS) combined with polarizing optical microscopy observations by Chen et al. demonstrated that the intermediate phase of DIO exhibits a lamellar type of order with a periodicity of 17.5 nm.[32] Within each period, there are two sublayers of a thickness $w =$ 8.8 nm each, manifested by an equilibrium sinusoidal electron-density modulation observed in non-resonant SAXS. The molecules are tangential to the layers and polarly ordered; the direction of polarity alternates from one subdomain to the next. The density modulation observed in the intermediate phase of DIO by Chen et al.[32] justifies the abbreviation "Sm", since it traditionally stands for a "smectic". "Z" reflects the fact that the average molecular orientation is parallel to the layers rather than perpendicular to them, as in a SmA. The subscript "A" stands for antiferroelectric. A different model was suggested recently by Rupnik et al.[33] on the basis of experimental studies of the intermediate phase in RM734, which in a pure material extends over a very narrow temperature range of about 1 °C. The N phase of RM734 shows a dramatic decrease of the splay elastic constant $K_1$ upon the approach to the $N_F$.[34] When an ionic fluid is added to RM734, the intermediate phase range increases dramatically and shows periodic domains of a period up to 10 µm. The observations are explained by the splay-modulated phase, abbreviated $N_S$, in which a reversal of the splay div **P** and of the average polarization **P** decreases the overall space charge and allows the structure to fill the space efficiently. In another study, a doubly splay-modulated antiferroelectric phase has been observed in RM734 in cells with ionic polymers as aligning layers.[35]

In what follows, we keep the abbreviation $SmZ_A$ for DIO since the material shows one-dimensional periodic density modulation, a defining property of a smectic order, directly revealed in SAXS,[32] and supported by light scattering experiments,[36] response to shear flow[18] and by the behavior of dislocations in DIO with chiral additives.[37] We use the same abbreviation for the intermediate antiferroelectric phase of FNLC919 since a similar density variation with a period of 6.7 nm has been documented in SAXS experiments by Paul et al.[38].

### 2.2 Rheometry

Rheological measurements are performed using a strain-controlled rheometer Anton Paar, MCR 302 with a parallel-plate measuring system having a plate diameter of 25 mm. The zero values of the normal force and torque were set for each test before introducing the sample. The sample was confined between the plates, and the excess material was removed. The plate gap was fixed to be 200 µm for RM734, 150 µm for DIO, and 100 µm for FNLC919; at these distances, the normal force was zero after the assembly of the test cells. No alignment layers were used. The top plate rotates at different shear rates while the bottom plate is fixed. A Peltier element was attached to the bottom plate to control the temperature with an accuracy of 0.1°C. The rheometer was placed under a hood to maintain temperature uniformity. The temperature dependence of effective shear viscosity was measured at a low shear rate of $\dot{\gamma} = 2.5$ s$^{-1}$ and a high shear rate of $\dot{\gamma} = 500$ s$^{-1}$ upon cooling and heating. The cooling and heating rates are 2 °/min in the range of 160-110 °C for RM734, 110-58 °C for DIO, and 85-18 °C for FNLC919, respectively.

### 2.3 In-situ polarizing optical microscopy

To characterize the shear-induced structural changes, we use the Linkam Optical Shearing System CSS450 in parallel disk geometry. The shear device features a window for observing textures using a polarizing optical microscope (POM) in the transmission mode. The bottom plate rotates relative to the top plate with a controllable angular velocity in the range (0.001-10) rad s$^{-1}$. Both plates contain quartz windows of 2.5 mm in diameter, centered at 6.25 mm from the axis of rotation. A separately controlled heater maintains the temperature of the sample fixed within 0.1 °C. No alignment layer is used so that the substrates of CSS450 yield degenerate tangential alignment of the director $\hat{\mathbf{n}}$. To reconstruct the structural response under shear flow, we use POM Olympus BX51 equipped with a video camera Baslar ac1920-155um (10-40 frames per second) with a full-wave-plate (FWP) 530 nm optical compensator. The slow axis $\boldsymbol{\lambda}_g$ of the optical compensator is oriented at 45° to the crossed polarizers.

## 2.4 Polychromatic polarizing microscopy

A polychromatic polarizing microscope (PPM) generates a map of director orientation in a single shot of a camera.[39] The PPM, invented by Shribak in 2009,[40] is equipped with a special polychromatic polarization state generator (PSG) and an achromatic circular analyzer. PPM produces a full hue-saturation-brightness (HSB) color spectrum in birefringent materials. The HSB hues in PPM depend on the slow axis orientation of the birefringent specimen with respect to a preset 'zero' direction (usually oriented along the East-West axis of a microscope's stage). This implies that the color changes occur continuously while the stage rotates, repeating every 180°, and the state of extinction is never observed. The PPM allows one to capture an image of fast-moving and low-birefringent structures in real-time, limited only by the acquisition time of the camera. The PPM approach is different from that of a conventional polarizing microscope, in which the retardance of the specimen determines the interference color in the Michel-Lavy chart.

The optical scheme of a PPM, which consists of a polychromatic PSG and an achromatic circular analyzer, with the N compensating cell, is shown in Fig. 2. The PSG produces polarized light with the polarization ellipse orientation that depends on the wavelength. PSG comprises a rotatable polarizer, an achromatic quarter-wave plate (AQWP), and an optically active waveplate (OAWP). The polarizer and AQWP produce a polarization ellipse with a major axis parallel to the slow axis of AQWP. When the elliptically polarized light travels along the optical axis of the waveplate OAWP, the polarization ellipse rotates by some angle. The rotation angle depends on the thickness of OAWP and the wavelength of light. An achromatic quarter-wave plate (AQWP) with a slow axis oriented at 45° and an analyzer are combined to form the achromatic left circular analyzer. The achromatic quarter-wave plate minimizes the variation in retardance across a broad spectral range.[41, 42]

A limitation of an original PPM device is that the recorded retardance of the birefringent specimen should be less than 250 nm. To overcome the limitation, we add an anisotropic optical compensator, representing a planar LC cell of the N material E7 (Jiangsu Hecheng Display Technology Co., Ltd. (HCCH), Jiangsu, China) with a slow axis (director) $\boldsymbol{\lambda}_N$. The N compensating cell introduces an optical retardance of an opposite sign to that of the sample, in order to reduce it below 250 nm. The N compensating cell is inserted between the sheared specimen and the achromatic left circular analyzer, with $\boldsymbol{\lambda}_N$ being perpendicular to the optic axis of the sheared material so that the retardance of this cell reduces the measured retardance of the system. The compensating cell is assembled from two flat glass substrates, spin-coated with a polyimide PI2555 (Nissan Chemicals, Ltd.), and rubbed unidirectionally to achieve a uniaxial planar alignment of $\boldsymbol{\lambda}_N$. The optical birefringence $\Delta n = 0.22$ of E7[43] (reported at the wavelength of 600 nm, room temperature $T = 25$ °C) is close to the birefringence of the explored materials, thus the thickness of the compensating cell was selected to be close to the thickness of the sheared samples. To generate the map of director orientations of the specimen from the HSB hue image of the PPM, spline interpolation of hue data has been performed using Mathematica code.[39]

## 2.5 Optical retardance and PolScope

To measure the optical retardance $\Gamma$ of the $N_F$ materials as a function of shear rate, we use the LC PolScope approach, invented by Oldenburg,[44, 45] and applications of which to liquid crystals have been described in Refs.[46, 47]. The LC PolScope represents a polarizing optical microscope with a variable optical compensator(s), which might be an N cell controlled by an applied electric field. Once the image of a sample is taken for a few different settings of the compensator, numerical analysis allows the unit to map the optical retardance $\Gamma$ of the sample and reconstructs the in-plane orientation of the optical axis $\hat{\mathbf{n}}$. The PolScope observations in this study are performed using the Exicor Microimager (Hinds Instruments), operating at four wavelengths: 475 nm, 535 nm, 615 nm, and 655 nm, which allows for the characterization of samples with optical retardance up to 3500 nm.

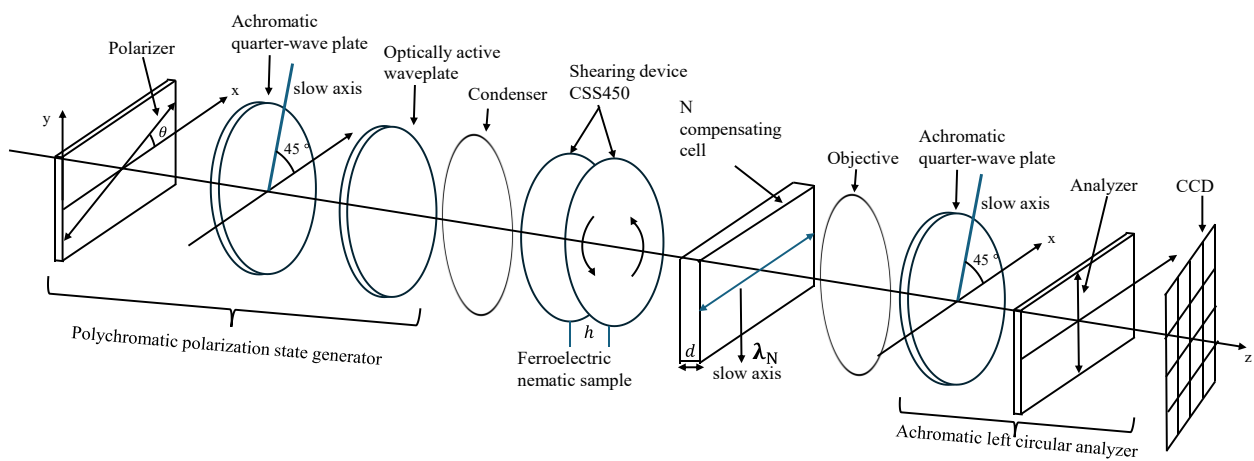

**Fig. 2**: Schematic of the experimental setup to characterize the structural changes of the ferroelectric nematic liquid crystal caused by shear flow using PPM with the N compensating cell. $\boldsymbol{\lambda}_N$ is the slow axis of the N compensating cell.

## 3. Results and discussion

### 3.1 Shear viscosity *vs.* temperature

The effective shear viscosity $\eta$ of all three materials in the N, $N_F$ and intermediate phases generally decreases with the temperature, Fig. 3a,b,c, a behavior typical for many N materials, including MBBA, 5CB and 8CB.[48-50] Notable features are pretransitional increases of $\eta$ and discontinuous changes at the phase transitions. In RM734, $\eta$ drops from 0.10 Pa·s in the $N_F$ phase at 115 °C to 0.03 Pa·s in the N phase at 160 °C; these values are similar to the previous results measured by Dhara et al.[17].

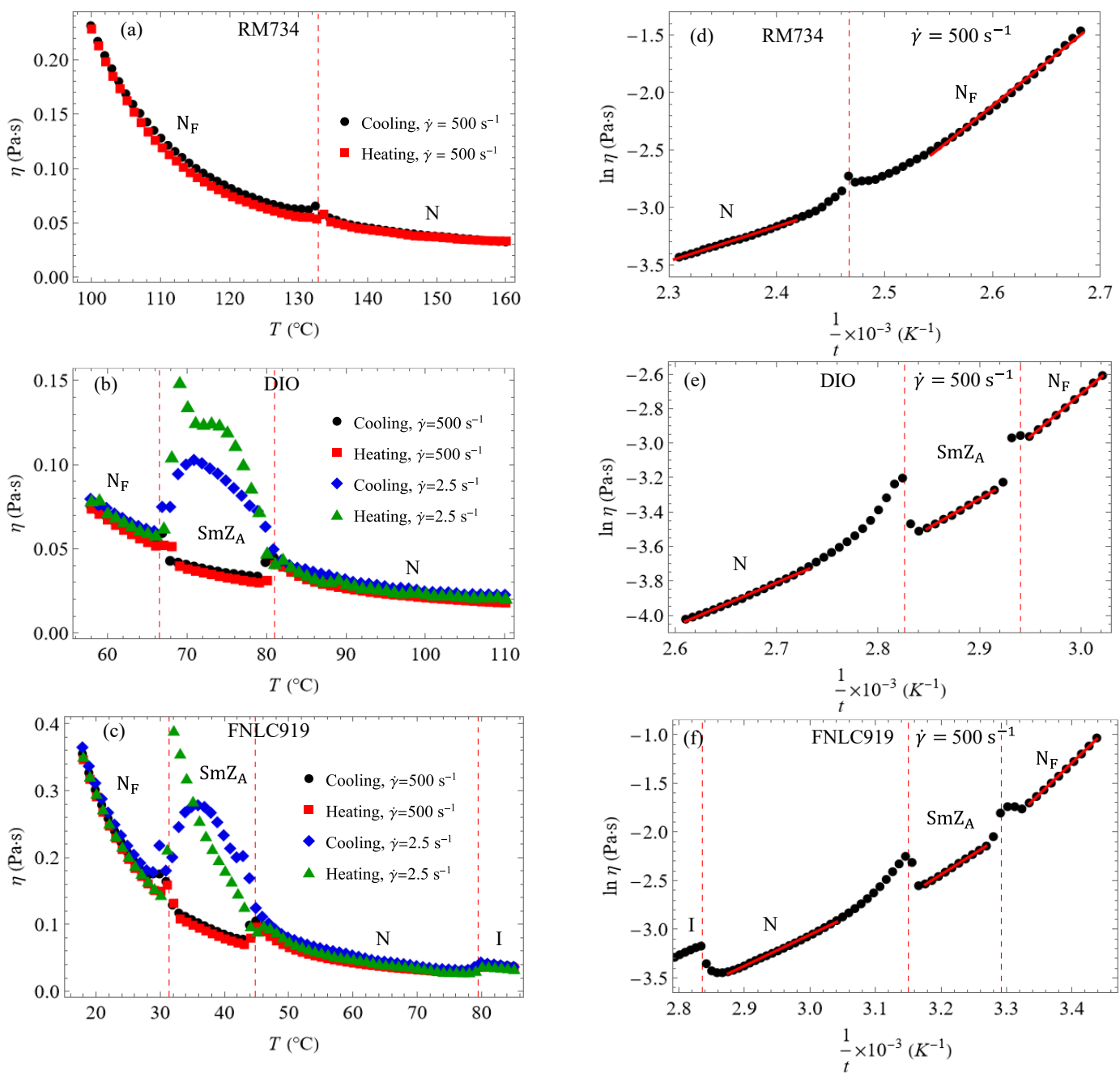

**Fig. 3**: The temperature dependence of the effective shear viscosity $\eta$ of RM734 (a) at a shear rate of $\dot{\gamma} = 500$ s$^{-1}$, DIO (b), and FNLC919 (c) at a shear rate of $\dot{\gamma} = 2.5$ s$^{-1}$ and $\dot{\gamma} = 500$ s$^{-1}$. The Arrhenius plot of the effective shear viscosity $\eta$ of RM734 (d), DIO (e), and FNLC919 (f) at the shear rate of $\dot{\gamma} = 500$ s$^{-1}$.

Table 1: Activation energy of RM734, DIO, and FNLC919 in the N, $N_F$, and intermediate phases

| RM734 | | DIO | | FNLC919 | |
|---|---|---|---|---|---|
| Phase, $T$ (°C) | $E_a$(kJ/mole) | Phase, $T$ (°C) | $E_a$(kJ/mole) | Phase, $T$ (°C) | $E_a$(kJ/mole) |
| N, 160 °C-140 °C | 25.7$\pm$0.2 | N, 110 °C-93 °C | 20.4$\pm$0.3 | N, 75 °C-55 °C | 25.3$\pm$0.3 |
| | | $SmZ_A$, 79 °C-70 °C | 27.8$\pm$0.2 | $SmZ_A$, 42 °C-33 °C | 34.9$\pm$0.2 |
| $N_F$, 120 °C-100 °C | 60.8$\pm$0.5 | $N_F$, 66 °C-58 °C | 41.1$\pm$0.3 | $N_F$, 28 °C-18 °C | 53.6$\pm$0.5 |

In the middle of the $N_F$ temperature range of RM734, the shear viscosity is $\eta = 0.13$ Pa·s, Fig. 3a. interestingly, this value is close to the rotational viscosity of RM734 $\gamma = 0.15$ Pa·s measured at the same temperature by rotating the polarization vector $\mathbf{P}$ by an electric field $\mathbf{E}$ applied perpendicularly to $\mathbf{P}$, which maximizes the realigning torque $\boldsymbol{\tau} = \mathbf{P} \times \mathbf{E}$.[51] Furthermore, the data for RM734, Fig. 3a, and DIO, Fig. 3b, can be compared to the effective viscosity $\bar{\eta}$, called "polarization reversal dissipation coefficient" and measured in the electro-optical responses of the $N_F$ phase to the field $\mathbf{E}$ that is antiparallel to $\mathbf{P}$.[52] Chen et al.[52] found this coefficient to be $\bar{\eta} = 0.05$ Pa·s at the highest temperature of the $N_F$ phase in both RM734 and DIO (and in all their binary mixtures), which is remarkably close to the values of shear viscosity $\eta = 0.06$ Pa·s in Fig. 3a,b at the highest temperatures of the $N_F$ phase. The polarization reversal dissipation coefficient $\bar{\eta}$ in Ref.[52] shows an Arrhenius-like temperature dependencies, which is again close to the behavior of the shear viscosity $\eta$, as discussed below.

The temperature dependencies of viscosity away from the transition points can be fitted by an Arrhenius law $\eta = \eta_0 e^{\frac{E_a}{k_B t}}$, Fig. 3d,e,f, with the activation energies $E_a$ listed in Table 1. Here, $k_B$ is the Boltzmann constant and $t$ is the absolute temperature. The activation energy $E_a$ in the $N_F$ phase is more than twice that in the N phase in all three materials. The viscosity of DIO and FNLC919 is almost the same at a low shear rate of $\dot{\gamma} = 2.5$ s$^{-1}$ and at a high shear rate of $\dot{\gamma} = 500$ s$^{-1}$ in the N and $N_F$ phases.

In contrast, the viscosity of the $SmZ_A$ in DIO and FNCL919 is significantly higher at a low shear rate of $\dot{\gamma} = 2.5$ s$^{-1}$ and lower at a high shear rate of $\dot{\gamma} = 500$ s$^{-1}$ than the viscosities of the neighboring regions of the N and $N_F$ phases. A similar behavior was reported for the $SmZ_A$ phase by Nishikawa et al.[18]. Smectic layers of the $SmZ_A$ between two parallel plates are oriented randomly at a low shear rate, with some of them being orthogonal to the flow, which would yield a high viscosity. At a high $\dot{\gamma}$, the layers are mostly parallel to the plates of the rheometer and slide over one another while keeping the average $\hat{\mathbf{n}}$ along the shear, which results in a lower viscosity. Note that this behavior is different from the behavior of the SmA phase in 8CB, in which $\hat{\mathbf{n}}$ is perpendicular to the layers and $\eta$ is higher than that of the N phase.[50, 53] The most viscous is the mixture FNLC919, while DIO is the least viscous, despite the fact that the temperature range of the DIO mesophases is lower than that of RM734.

**3.2 Shear viscosity *vs.* shear rate**

The variation of effective shear viscosity $\eta$ of all three materials with the shear rate in the range 0.1 s$^{-1} \leq \dot{\gamma} \leq 1000$ s$^{-1}$ has been explored in the N, $N_F$ and intermediate phases, Fig. 4. All three materials and all three phases show a pronounced shear-thinning at low $\dot{\gamma} < 1$ s$^{-1}$ and almost Newtonian behavior at high $\dot{\gamma} > 100$ s$^{-1}$. The intermediate range $1$ s$^{-1} < \dot{\gamma} < 100$ s$^{-1}$ exhibits a moderate shear-thinning. The shear rate dependencies can be presented by a power law $\eta \propto \dot{\gamma}^{n-1}$, in which $n < 1$ describes a shear-thinning and $n = 1$ corresponds to the Newtonian flow behavior.

The measured viscosities of the DIO in the $N_F$ phase are higher than the one reported previously for a similar range of shear rates in Ref.[18]. For example, at 60 °C, $\eta = (0.07 - 0.09)$ Pa·s in Fig. 3b, Fig. 4b for $\dot{\gamma} = (2.5 - 1000)$ s$^{-1}$, while $\eta = 0.04$ Pa·s in Fig. 4d of Ref.[18], measured at $\dot{\gamma} = 2.6$ s$^{-1}$ and 5000 s$^{-1}$. Another notable difference is that all three phases of DIO in Fig. 4b show a shear-thinning regime and almost a Newtonian behavior at high shear rates, while Fig. 3 in Ref.[18], reports a change from shear-thinning to shear-thickening regime in the $N_F$ phase of DIO at $\dot{\gamma} > 20$ s$^{-1}$, although with an exponent $n = 1.02$ close to 1. A potential reason for these discrepancies is the different thickness of shear cell, 150 µm in our case and 80 µm in Ref.[18].

**3.3 First-normal stress difference *vs.* shear rate**

The variation of the first-normal stress difference $N_1$ of all three materials as a function of shear rate in the range of 0.1 s$^{-1} \leq \dot{\gamma} \leq 1000$ s$^{-1}$ has been observed in the N, $N_F$ and intermediate phases, Fig. 5. $N_1$ is the difference between the normal stress along the flow and the normal stress in the orthogonal direction of the velocity gradient. $N_1$ is a measure of the non-Newtonian and elastic behavior of fluids under shear; in Newtonian isotropic fluids, $N_1 = 0$. As a rule, isotropic fluids such as polymer solutions, show $N_1 > 0$. The reason is that the polymer coils extend along the flow, which creates a restoring force that tends to return the coil to the initial isotropic shape, thus acting to push the plates of the shear device apart, hence $N_1 > 0$. It thus came as a surprise that some polymers with orientational order show $N_1 < 0$.[54-57] The qualitative explanation is that in these polymers, a moderate shear causes tumbling of the director $\hat{\mathbf{n}}$, which results in a less ordered structure than that in the absence of shear. All three explored materials show $N_1 < 0$ in the N and intermediate phases. The $N_F$ phase in all materials exhibits a small negative $N_1$ at low shear rates $\dot\gamma < 10\,\mathrm{s}^{-1}$ and a positive $N_1$ at $\dot\gamma > 100\,\mathrm{s}^{-1}$. The experiment uncovers a dramatic difference in the behavior of $N_1$ between the paraelectric, antiferroelectric and ferroelectric phases. The observed behavior does not fit fully the previously developed models. The available models that explain $N_1 < 0$ [58,59] in the N phase assume that $\hat{\mathbf{n}}$ is in the shear plane; as will be clear in the next section, this assumption is valid only at $\dot\gamma \leq 100\,\mathrm{s}^{-1}$; at higher rates, polydomain structures form and at $\dot\gamma \geq 1000\,\mathrm{s}^{-1}$, $\hat{\mathbf{n}}$ realigns along the vorticity axis. When $\hat{\mathbf{n}}$ deviates from the shear plane, forming twisted structures,[60] the available models predict $N_1 > 0$.[61,62]

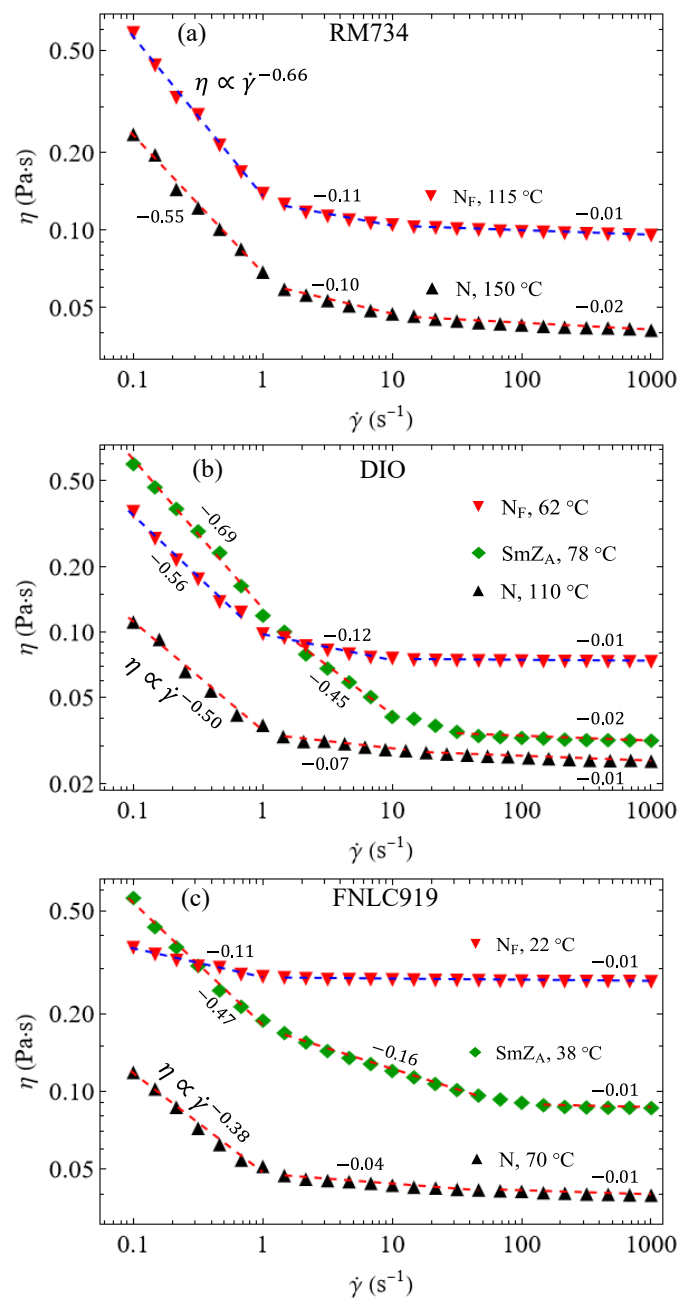

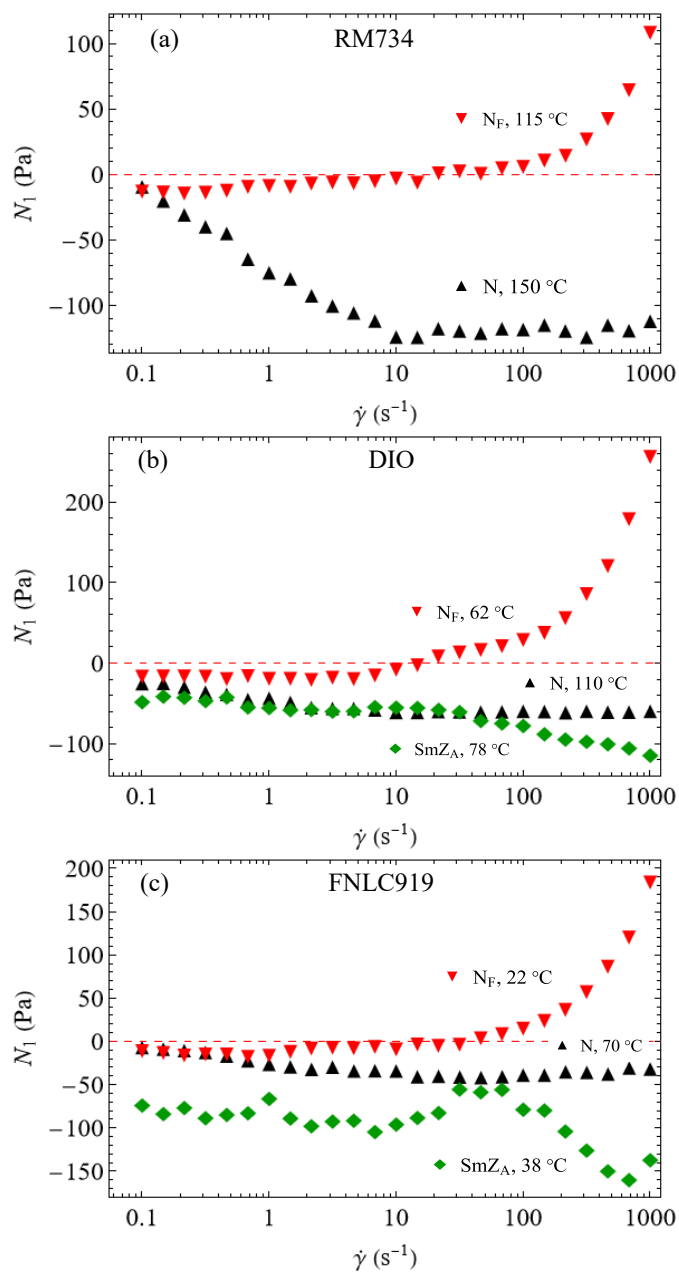

**Fig. 4**: Variation of effective shear viscosity $\eta$ of RM734 (a), DIO (b), and FNLC919 (c) with different shear rates $\dot\gamma$ in the N and $N_F$ and intermediate phases. The red and blue dashed lines are the best fit to the curve, which is done by using a power law for effective shear viscosity.

**Fig. 5**: First normal stress difference $N_1$ of RM734 (a), DIO (b), and FNLC919 (c) as a function of shear rate $\dot\gamma$ in the N, $N_F$ and intermediate phases.

For the case of the $N_F$ phase, a small negative $N_1$ at low $\dot\gamma$ can be tentatively attributed to misalignments in the shear plane, but the optical retardance discussed below does not support this idea. At high rates, a positive $N_1$ can be associated with a better alignment of the director as compared to shear-free case; however, there is no significant difference in the optical retardance of the samples in these two regimes. Intriguing behavior of $N_1$ deserves further studies.

### 3.4 Realignment regimes of the N phase under shear
*RM734.* The response of the N phase at $T = 150\,°\mathrm{C}$ to the shear exhibits three regimes, depending on the shear rate. As $\dot\gamma$ increases from 0.75 $s^{-1}$ to 80 $s^{-1}$, initially misaligned $\hat{n}$ progressively realigns toward the shear plane, as evidenced by PPM, Fig. 6a and by observations with the optical compensator, Supplementary Fig. S1. The realignment of $\hat{n}$ towards the shear plane indicates a flow-aligning character of RM734 in the N phase at low $\dot{\gamma}$. Similar behavior is observed in conventional N materials, such as MBBA, 5CB, E7, and MLC 7026.[13,14,63,64] As $\hat{\mathbf{n}}$ remains in the shear plane, the elastic deformations are predominantly of splay-bend type. The relative importance of the viscous and elastic torques is expressed by the Ericksen number $\mathrm{Er} = \eta\dot{\gamma}h^2/K_1$, where $K_1$ is the splay elastic constant. Using typical values $K_1 \approx 2$ pN at 150 °C,[34] cell thickness $h = 10^{-5}$ m, the effective viscosity $\eta = 0.04$ Pa·s at 150 °C, Fig. 3a, one estimates $\mathrm{Er} \approx [2\ \mathrm{s}]\ \dot{\gamma}$. In other words, the viscous torques prevail over the elastic ones at any $\dot{\gamma} > 0.5\ \mathrm{s}^{-1}$. As the shear rate increases to $\dot{\gamma} = 120\ \mathrm{s}^{-1}$, flow starts to produce disclination loops that multiply with the further growth of $\dot{\gamma}$, Fig. 6b, Supplementary Fig. S2a. In the range $120\ \mathrm{s}^{-1} \leq \dot{\gamma} \leq 1000\ \mathrm{s}^{-1}$, the director field is highly distorted, forming disclinations-infused polydomains which become more elongated as $\dot{\gamma}$ increases, Fig. 6b, Supplementary Fig. S2a. The director adopts many different orientations in the plane of the sample. At still higher shear rates, $1000\ \mathrm{s}^{-1} \leq \dot{\gamma} \leq 5000\ \mathrm{s}^{-1}$, the polydomain textures become progressively homogeneous with the director $\hat{\mathbf{n}}$ gradually realigning towards the vorticity $x$-axis, as evidenced by the predominance of yellow colors in textural observation with an optical compensator in Fig. 6c and blue colors in Supplementary Fig. S2b. This regime can be called log-rolling.

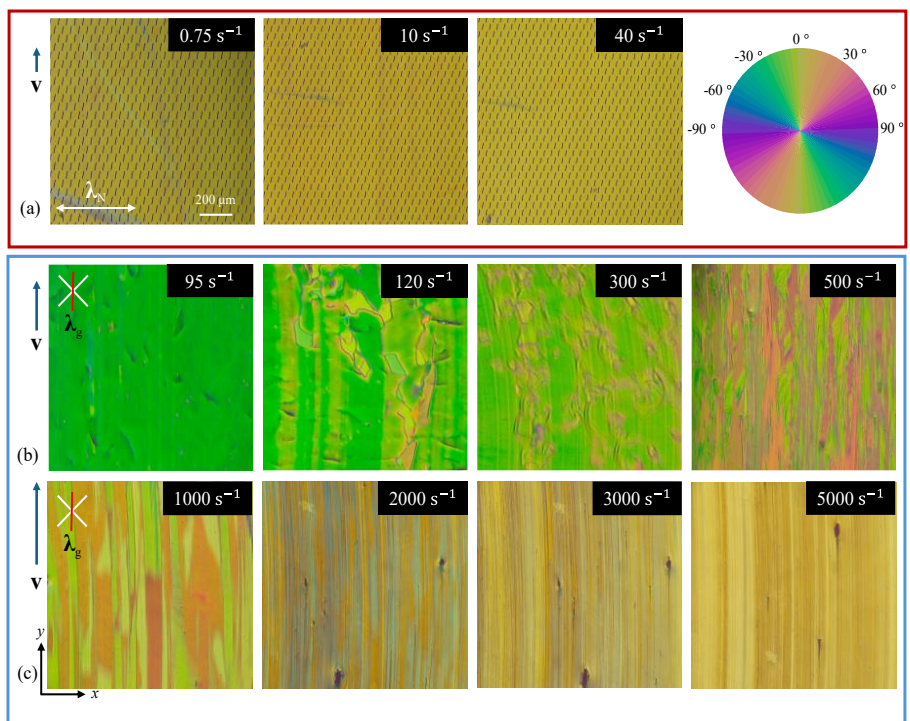

**Fig. 6**: Structural response of the N phase of RM734 to shear at $T = 150$ °C in a 10 μm thick cell revealed in the PPM (a) and POM (b,c) modes of observation; the shear rate $\dot{\gamma}$ is indicated on the textures. (a) Flow-alignment at low $\dot{\gamma}$ as revealed by PPM with the N compensating cell of a thickness 9.6 μm. The color wheel describes the orientation of the optic axis (director) and the ticks in textures show the local orientation of the director. (b) Polydomain textures at the intermediate $\dot{\gamma}$ as revealed by POM with crossed polarizers and a FWP compensator. The slow axis $\lambda_g$ of the FWP shown by a red line is parallel to the shear direction. (c) Progressive realignment of the director towards the vorticity direction at high $\dot{\gamma}$. The slow axis $\lambda_g$ of the FWP is parallel to the shear direction.

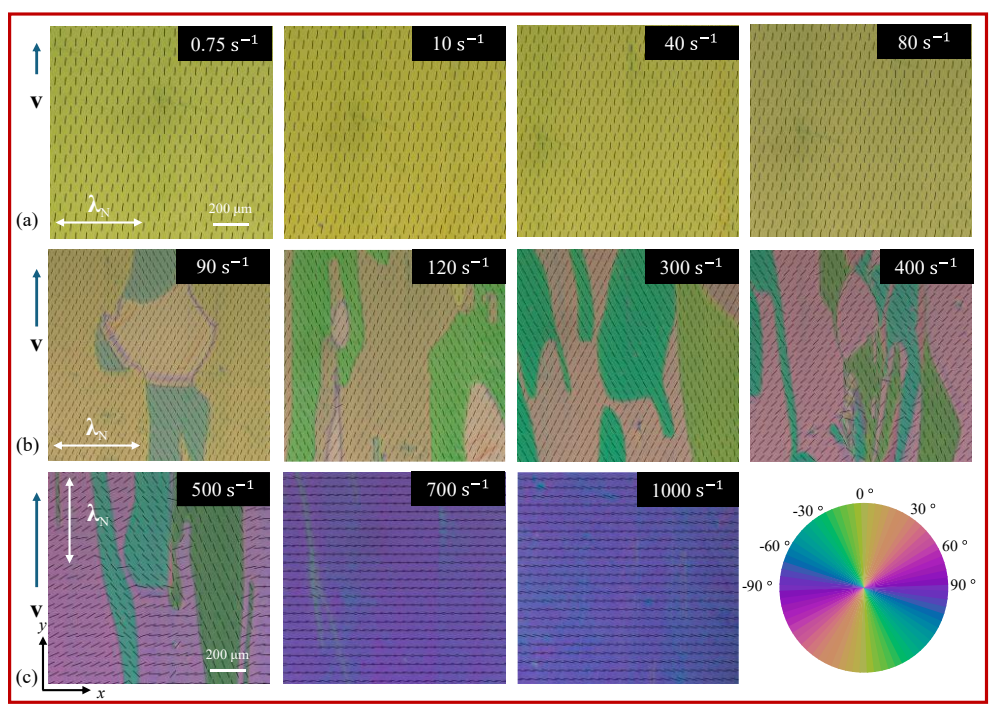

**Fig. 7**: Structural response of the N phase of DIO to shear at $T = 110$ °C in a 10 μm thick cell revealed in the PPM modes of observation; the shear rate $\dot{\gamma}$ is indicated on the textures. (a) Flow-alignment at low $\dot{\gamma}$ as revealed by PPM with the N compensating cell of a thickness 6.8 μm. The color wheel describes the orientation of the optic axis (director) and the ticks in textures show the local orientation of the director. (b) Polydomain textures at the intermediate $\dot{\gamma}$. (c) Progressive realignment of the director towards the vorticity direction at high $\dot{\gamma}$.

_DIO and FNLC919._ The N phase of DIO and FNLC919 exhibits three similar flow regimes. The director, which is initially misaligned, realigns along the shear plane at low shear rates, $0.75\ \mathrm{s}^{-1} \leq \dot{\gamma} \leq 80\ \mathrm{s}^{-1}$ in DIO and $0.75\ \mathrm{s}^{-1} \leq \dot{\gamma} \leq 50\ \mathrm{s}^{-1}$ in FNLC919, Fig. 7a, Supplementary Fig. S3a, S4, S5. Intermediate rates bring about polydomain structures, Fig. 7b, Fig. S3b. The highest shear rates, $600\ \mathrm{s}^{-1} \leq \dot{\gamma} \leq 1000\ \mathrm{s}^{-1}$, reorient $\hat{\mathbf{n}}$ along the vorticity $x$-axis, Fig. 7c, Supplementary Fig. S3c,d, producing a log-rolling regime.

The observed realignment of the N molecules along the vorticity direction is at odds with the previously reported flow-alignment of the DIO N phase in the shear plane at $\dot{\gamma} > 10^2\ \mathrm{s}^{-1}$.[18] Note that the structural analysis in Ref.[18] was performed by observations between crossed polarizers; in such a setting, it is difficult to distinguish between two orthogonal directions of the optic axis. Such an ambiguity is removed when the observations are performed with optical compensators, Fig. 6b,c and in the PPM mode, Fig. 7.

### 3.5 Realignment regimes of the $N_F$ phase under shear

_RM734_ The $N_F$ realignment under shear is qualitatively similar to that of the N phase, Fig. 8. At low shear rates $\dot{\gamma} \leq 0.75\ \mathrm{s}^{-1}$, the flows are not strong enough to streamline polydomain textures. When $\dot{\gamma}$ increases from 0.75 $\mathrm{s}^{-1}$ to 5 $\mathrm{s}^{-1}$, the polydomain texture slowly (within ~10 min) transforms into a homogeneous texture with $\hat{\mathbf{n}}$ approaching the shear plane, Fig. 8a. The $N_F$ behaves as a flow-aligning material in the range of 5 $\mathrm{s}^{-1} \leq \dot{\gamma} \leq 40\ \mathrm{s}^{-1}$. Above this range, $45\ \mathrm{s}^{-1} \leq \dot{\gamma} \leq 300\ \mathrm{s}^{-1}$, the flow creates a polydomain structure, Fig. 8b,c. At $\dot{\gamma} > 300\ \mathrm{s}^{-1}$, the shear progressively realigns the optic axis towards the vorticity $x$-axis, Fig. 8c. The regime is log-rolling when $700\ \mathrm{s}^{-1} \leq \dot{\gamma} \leq 1000\ \mathrm{s}^{-1}$, Fig. 8c.

_DIO and FNLC919._ The flow aligns $\hat{\mathbf{n}}$ of the $N_F$ in the shear plane at $3\ \mathrm{s}^{-1} \leq \dot{\gamma} \leq 30\ \mathrm{s}^{-1}$, Fig. 9a, Fig. S6a. The intermediate range, $35 \text{ s}^{-1} \leq \dot{\gamma} \leq 350 \text{ s}^{-1}$ in DIO and $35 \text{ s}^{-1} \leq \dot{\gamma} \leq 260 \text{ s}^{-1}$ in FNLC919, produces polydomain textures; the number of domains increases with $\dot{\gamma}$, Fig. 9b, Fig. S6b. The director progressively turns towards the vorticity direction at $\dot{\gamma} > 350 \text{ s}^{-1}$ in DIO and $\dot{\gamma} > 260 \text{ s}^{-1}$ in FNLC919, Fig. 9c, Fig. S6c. At high shear rates $500 \text{ s}^{-1} \leq \dot{\gamma} \leq 1000 \text{ s}^{-1}$, $\hat{\mathbf{n}}$ is along the vorticity $x$-axis, Fig. 9c, Fig. S6c.

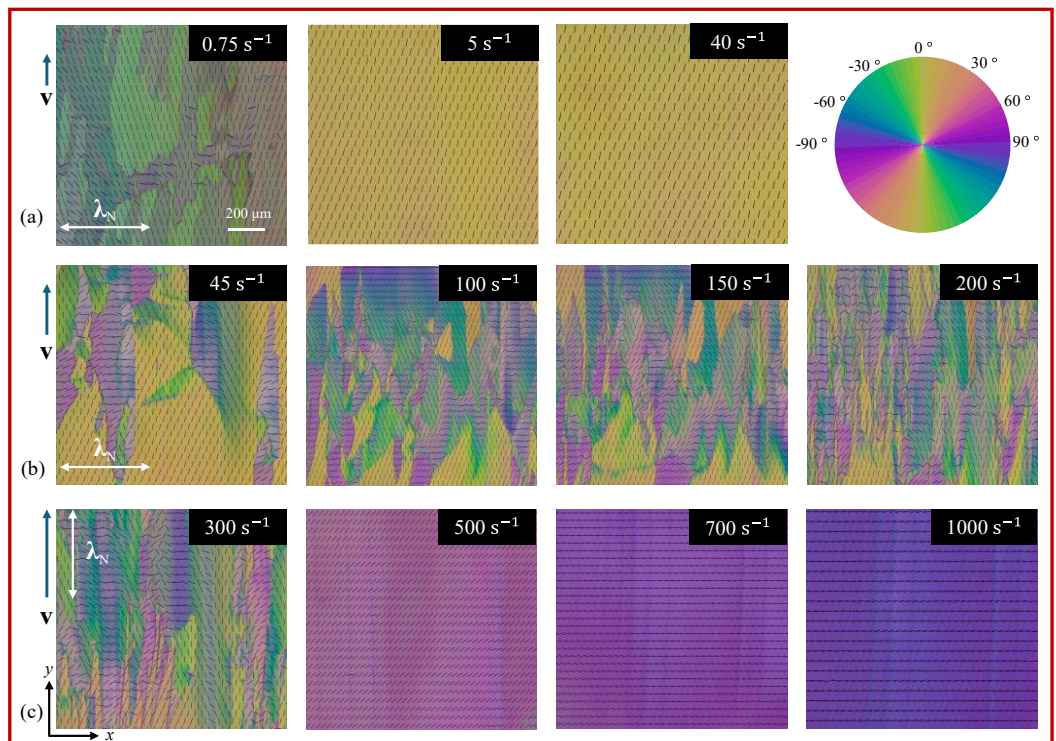

**Fig. 8**: Structural response of the $N_F$ phase of RM734 to shear at $T = 125$ °C in a 10 μm thick cell revealed in the PPM modes of observation; the shear rate $\dot{\gamma}$ is indicated on the textures. (a) Flow-alignment at low $\dot{\gamma}$ as revealed by PPM with the N compensating cell of a thickness 11.1 μm. The color wheel describes the orientation of the optic axis (director) and the ticks in textures show the local orientation of the director. (b) Polydomain textures at the intermediate $\dot{\gamma}$. (c) Progressive realignment of the director towards the vorticity direction at high $\dot{\gamma}$.

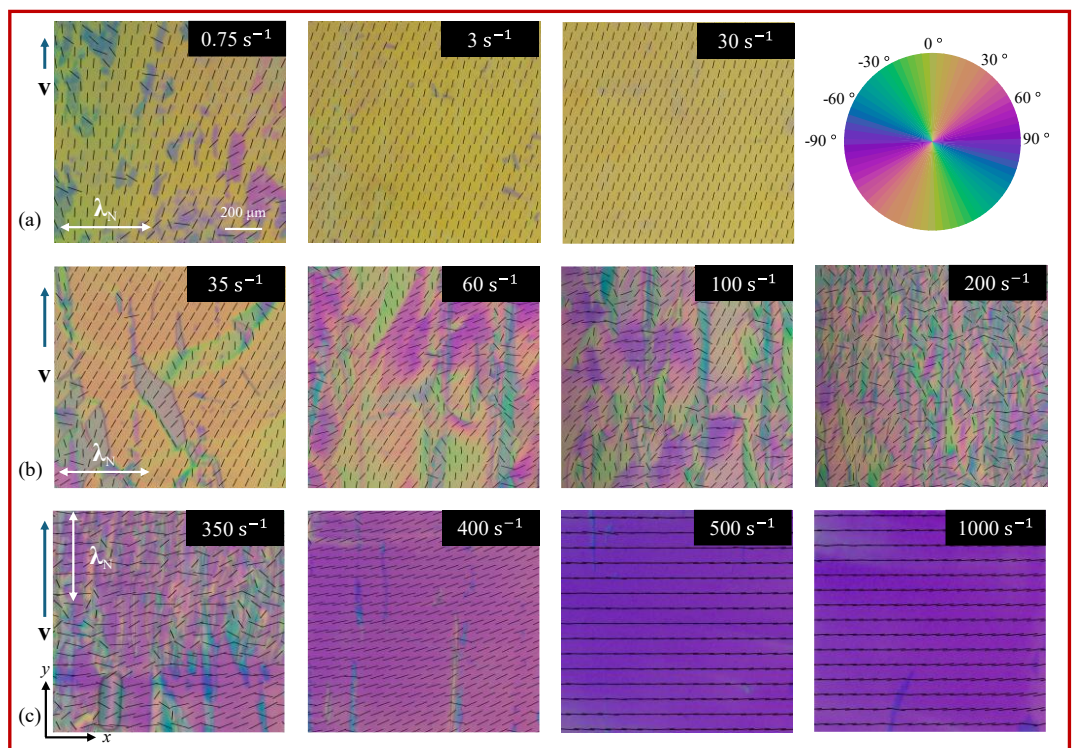

**Fig. 9**: Structural response of the $N_F$ phase of DIO to shear at $T = 65$ °C in a 10 μm thick cell revealed in the PPM modes of observation; the shear rate $\dot{\gamma}$ is indicated on the textures. (a) Flow-alignment at low $\dot{\gamma}$ as revealed by PPM with the N compensating cell of a thickness 8.4 μm. The color wheel describes the orientation of the optic axis (director) and the ticks in textures show the local orientation of the director. (b) Polydomain textures at the intermediate $\dot{\gamma}$. (c) Progressive realignment of the director towards the vorticity direction at high $\dot{\gamma}$.

The observed realignment of the $N_F$ molecules along the vorticity direction contradicts the previously reported flow-alignment of the DIO $N_F$ phase in the shear plane at $\dot{\gamma} > 10^2 \text{ s}^{-1}$.[18] As mentioned before, the textures in Fig. 8,9, recorded in the PPM mode, allow one to unambiguously determine the direction of the optic axis and polarization along the vorticity direction.

### 3.6 Director orientation under shear

To quantify the effect of flows on the director structures, we measure the optical retardance $\Gamma$ as a function of shear rate in the N and $N_F$ phases, Fig. 10a,b,c. In a separate experiment, we measured the retardance $\Gamma_{max} = \Delta n \times h$ of planar cells of the same thickness $h = 10$ μm as the rheometer gap, in the absence of shear; $\Delta n = n_e - n_o$ is the birefringence of material, $n_e$ and $n_o$ are the extraordinary and ordinary refractive indices, respectively. The $\Gamma_{max}$ values are shown as dashed lines in Fig. 10a,b,c. In the N phase of RM734 at 150 °C, $\Delta n = 0.195$ (measured at $\lambda = 655$ nm)[65] and $\Gamma_{max} = 1950$ nm. The N phase of DIO and FNLC919 show $\Gamma_{max} = 1540$ nm at 110 °C and 1500 nm at 65 °C, respectively ($\lambda = 655$ nm), Fig. 11a,b. These values are assumed as corresponding to the director parallel to the bounding plates; the effects of a small (on the order of 1°) pretilt angle are neglected.

*N phase.* In the N phase of all materials, at low shear rates, $\dot{\gamma} \leq 100 \text{ s}^{-1}$, $\Gamma < \Gamma_{max}$, Fig. 10a,b,c. The result is natural for a flow-aligning behavior since $\hat{\mathbf{n}}$ in the shear plane tilts away from the horizontal flow direction by some angle $\theta$. For example, in the flow-aligning nematics MBBA and 5CB, $\theta = (7 - 15)°$.[66-68] When $\hat{\mathbf{n}}$ makes an angle $\theta(z)$ with the y-axis in the flow-alignment regime, $\Gamma = \int_{z=0}^{z=h} \left( \frac{n_e n_o}{\sqrt{n_e^2 \sin^2\theta(z) + n_o^2 \cos^2\theta(z)}} - n_o \right) dz$. In a flow-aligning N, it is safe to assume that $\theta(z) = \theta_0$ everywhere except in the thin subsurface layers where the anchoring-imposed orientation persists.[66-70] We obtained $n_e$ and $n_o$ for RM734 from Ref.[16] and measured the values for DIO and FNLC919, Fig. 11c,d, using a wedge-cell interference technique.[71] Using the values of $h$, $n_e$ and $n_o$, we determine $\theta_0$ to be in the range (10-15)° in the flow-aligning regime, Fig. 10d,e,f. These results are close to what was previously measured in flow aligning MBBA and 5CB.[66, 67, 69]

In the range $100 \text{ s}^{-1} \leq \dot{\gamma} \leq 2000 \text{ s}^{-1}$ for RM734 and $100 \text{ s}^{-1} \leq \dot{\gamma} \leq 500 \text{ s}^{-1}$ for DIO and FNLC919, the polydomain texture yields only some effective $\Gamma$ since the director field is strongly distorted; twist and light scattering at the defects, Fig. 6b, Supplementary Fig. 2a, also diminish reliability of the $\Gamma$ data. At very high shear rate, $\dot{\gamma} > 2000 \text{ s}^{-1}$ for RM734 and $\dot{\gamma} > 500 \text{ s}^{-1}$ for DIO and FNLC919, when $\hat{\mathbf{n}}$ is along the vorticity axis, $\Gamma$ becomes practically equal to $\Gamma_{max}$, Fig. 10a,b,c. There is thus no significant departure of $\hat{\mathbf{n}}$ from the planar state in the log-rolling regime.

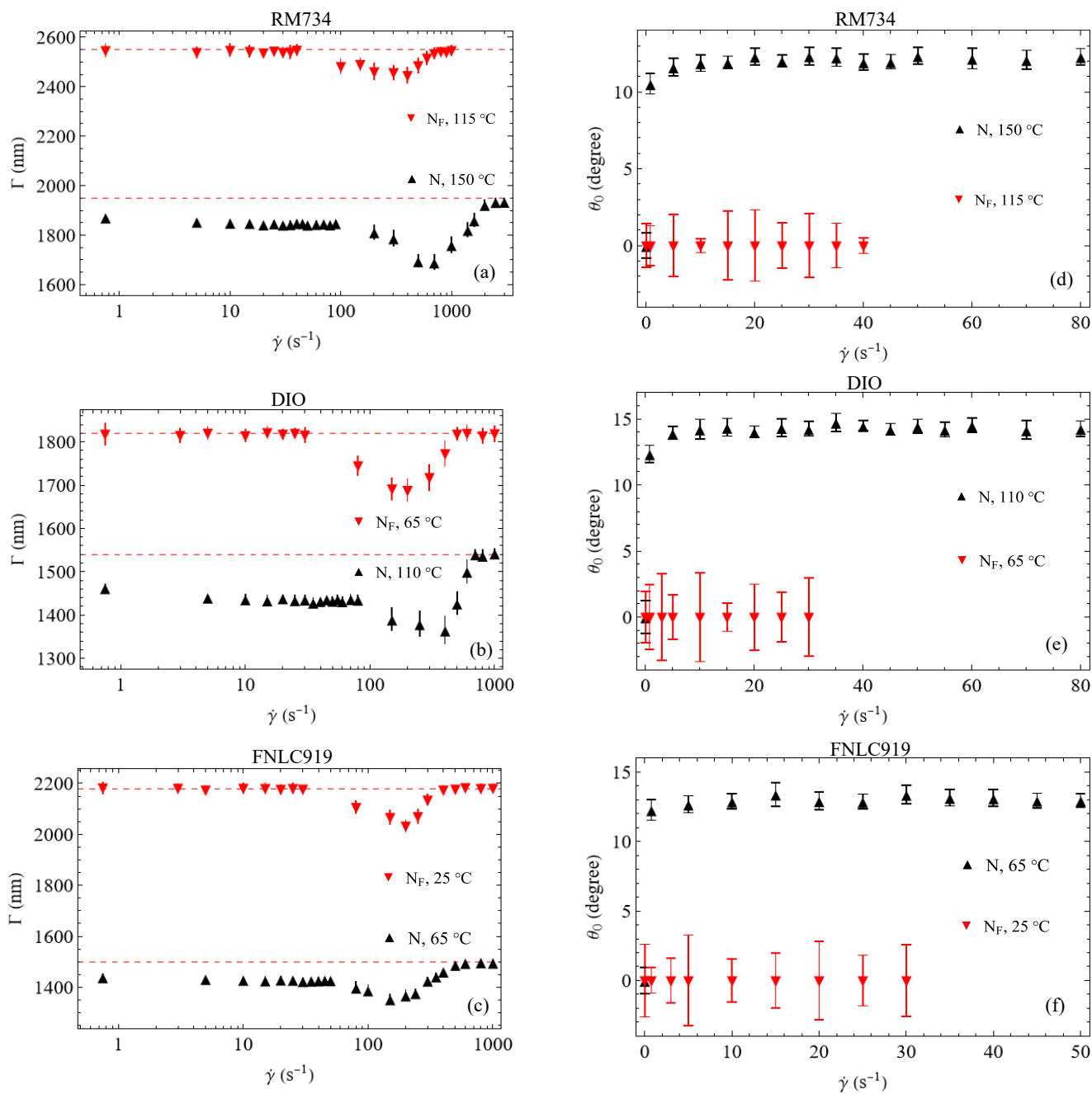

**Fig. 10**: Shear rate dependence of effective retardance $\Gamma$ measured by PolScope Microimager for RM734 (a), DIO (b), FNLC919 (c), and flow-alignment angle $\theta_0$ of RM734 (d), DIO (e), and FNLC919 (f) in the N and $N_F$ phases; error bars are standard deviations. The dashed red lines indicate the maximum possible retardance $\Gamma_{max}$ of a planar cell of 10-μm thickness in the absence of shear. $\Gamma$ of RM734 is measured at $\lambda = 655$ nm in the N phase and $\lambda = 535$ nm in the $N_F$ phase. $\Gamma$ of DIO and FNLC919 are measured at $\lambda = 655$ nm in the N and $N_F$ phases.

<u>$N_F$ phase.</u> At low shear rates, in the flow-aligning regime, all three materials in the $N_F$ phase show $\Gamma = \Gamma_{max}$, Fig. 10a,b,c. For example, for RM734, $\Gamma = \Gamma_{max} \approx 2550$ nm (at $\lambda = 535$ nm) for a 10 μm thick homogeneous planar sample at 125 °C.[65] It means that the optical axis (and thus the polarization **P**) does not deviate from the flow direction, in a stark contrast to the behavior in the N. The tilt of **P** at the bounding surfaces and splay of **P** in the bulk create bound charges and increases the electrostatic energy.[4, 72] This electrostatic mechanism explains the difference in the flow-alignment of the $N_F$ and N. At very high shear rates, in the log-rolling regime, one also observes $\Gamma = \Gamma_{max}$, which means that **P** avoids tilts and splay, but is now oriented along the vorticity $x$-axis. At intermediate shear rates, the retardance of the $N_F$ phase is much lower than $\Gamma_{max}$, Fig. 10a,b,c. The reason is the strong director distortions, with a prominent presence of twist along the velocity gradient $z$-axis. The twist is evident in POM observations with decrossed polarizers, Fig. 12: regions 1, 2, and 3 show different colors when the analyzer is rotated clockwise, Fig. 12a, and then

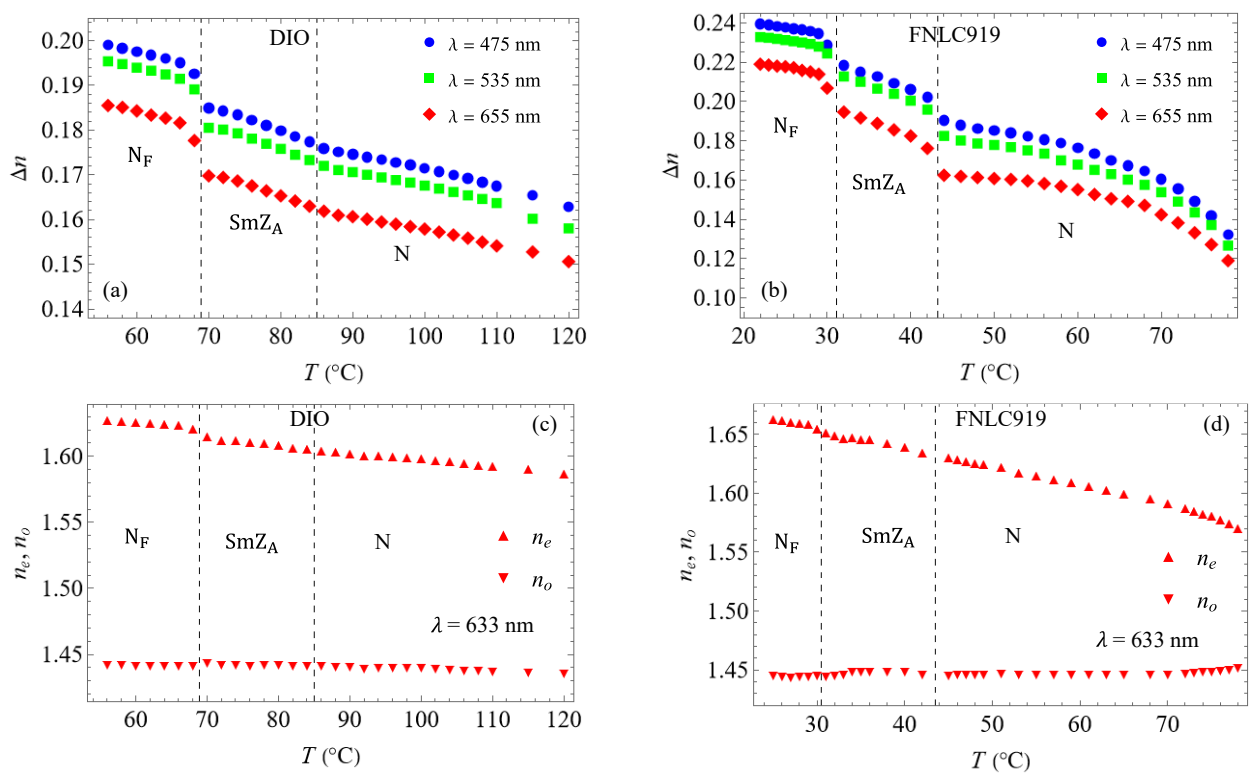

Fig. 11: Temperature dependence of birefringence $\Delta n$ of DIO (a) and FNLC919 (b) measured by PolScope Microimager at wavelengths 475 nm, 535 nm, 655 nm, and temperature dependence of the refractive indices of DIO (c) and FNLC 919 (d) measured by an interference technique using a wedge cell at wavelength 633 nm.

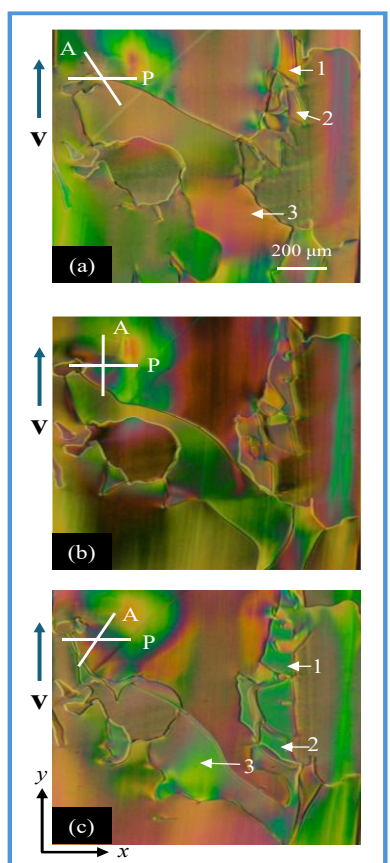

Fig. 12: Twisted domains of the sheared sample RM734 in the $N_F$ phase at $\dot\gamma = 100$ s$^{-1}$. [(a)–(c)] POM textures with the polarizer and analyzer making angles of 120°, 90°, and 60°, respectively. 10 μm thick sample and $T = 125$ °C.

quickly (0.5 s) rotated counterclockwise by the same angle, Fig. 12c.

## Conclusions

We performed comparative analysis of the paraelectric N, ferroelectric $N_F$, and antiferroelectric intermediate phases in three liquid crystal materials, RM734, DIO and FNLC919. As expected, the effective shear viscosity in all phases increases as the temperature decreases, demonstrating the Arrhenius's behavior except near the phase transition temperature. All three materials exhibit a strong shear-thinning behavior at low shear rates, $0.1$ s$^{-1} \leq \dot\gamma < 10$ s$^{-1}$ and a nearly Newtonian flow behavior at $\dot\gamma > 100$ s$^{-1}$. Shear-thinning is especially pronounced in the antiferroelectric phase, which is caused by its layered structure and progressive alignment of the layers under the shear. The same feature produces a dramatic difference in the effective viscosity measured at a constant temperature for two different shear rates: a low shear rate, 2.5 s$^{-1}$, does not align the layers well and the effective viscosity is very high, while high shear rate of 500 s$^{-1}$ aligns the layers well and the effective viscosity becomes even lower than that of the N and $N_F$ phases, Fig. 3b,c.

The first-normal stress difference $N_1$ shows an intriguing behavior, being negative in the N and intermediate phases, but changing from a small negative to a large positive value in the $N_F$ phase as $\dot\gamma$ increases. The behavior does not fit the available models developed for the N phase.

The structural response to the shear in both N and $N_F$ shows three distinct regimes. (I) Flow-alignment at low shear rates $\dot\gamma$

$< 10^2$ s$^{-1}$, with the director in the shear plane, making an angle $(10-15)°$ with the flow direction in the N and $0°$ in the N$_F$. (II) Polydomain textures with strong director deformations, including twists, at intermediate shear rates. (III) Log-rolling at high shear rates $\dot{\gamma} > 10^3$ s$^{-1}$, in which the director in the N and the polarization **P** in the N$_F$ are parallel to the vorticity direction. The absence of tilts and splay deformations in the flow-aligning and log-rolling regimes is rooted in the electrostatic properties of the N$_F$ phase, which avoids creation of surface and bulk space charges. The uncovered rheological properties and structural dynamics under shear would be useful for a better understanding of the N$_F$ materials and their potential applications in microfluidic devices.

## Author contributions

A.C.D. performed the experiments, analyzed the data, and wrote the manuscript. S.P. performed the refractive index measurements, analyzed the results, and contributed to writing that section. O.D.L. conceived the idea, supervised the overall research, and contributed to writing the manuscript.

## Conflicts of interest

There are no conflicts to declare.

## Data availability

All data needed to evaluate the conclusions in the paper are present in the article and in the Supplementary Figures. The datasets generated during and/or analyzed during this study are available from the corresponding author on request.

## Acknowledgements

The authors thank Merck KGaA, Darmstadt, Germany for providing the material FNLC919 and Dr. Hari Krishna Bisoyi and Organic Synthesis Facility at the AMLCI for the synthesis and purification of DIO. The work was supported by NSF grant DMR-2341830.

# Rheological properties and shear-induced structures of ferroelectric nematic liquid crystals

Ashish Chandra Das,[a,b] Sathyanarayana Paladugu,[b] and Oleg D. Lavrentovich *[a,b,c]

[a]Materials Science Graduate Program, Kent State University, Kent, OH 44242, USA

[b]Advanced Materials and Liquid Crystal Institute, Kent State University, Kent, OH 44242, USA

[c]Department of Physics, Kent State University, Kent, OH 44242, USA

**Supplementary Information**

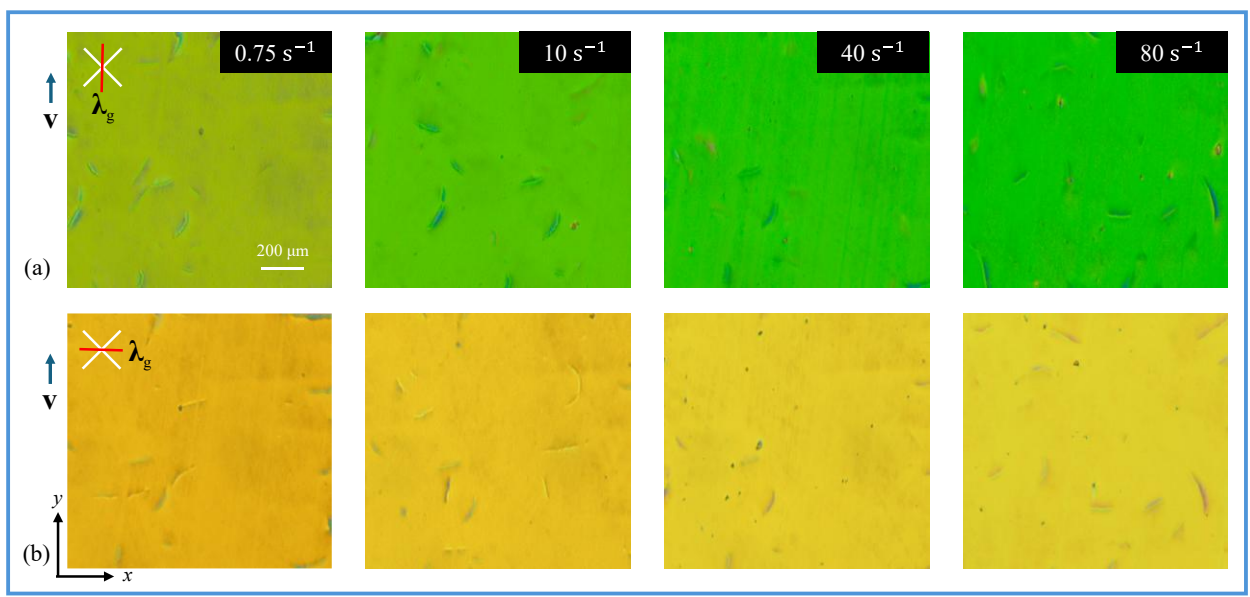

**Supplementary Fig. S1**: Structural change of RM734 with an increasing shear rate in the N phase, as revealed by POM with crossed polarizers and FWP's slow axis $\lambda_g$ aligned (a) parallel to the shear direction, and (b) perpendicular to it; a 10 μm thick sample and $T = 150$ °C. Shear realigns the director toward the shear plane.

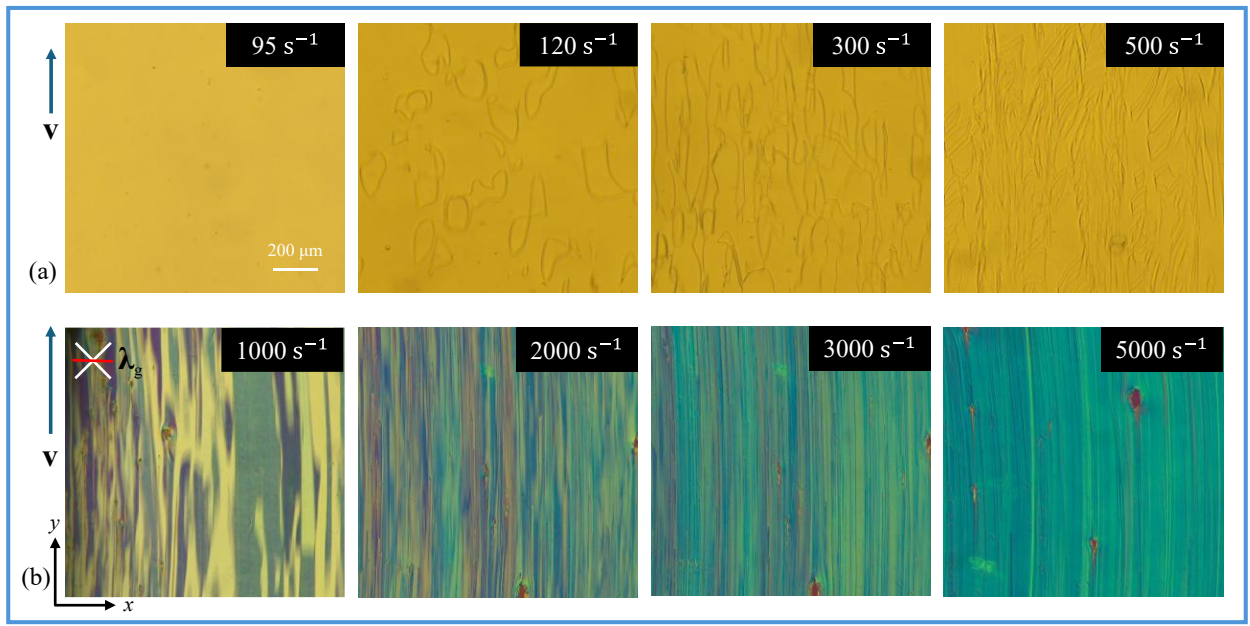

**Supplementary Fig. S2**: Structural response of the N phase of RM734 to shear at $T = 150\ °C$ in a 10 μm thick cell revealed in the POM (a,b) mode of observation; the shear rate $\dot{\gamma}$ is indicated on the textures (a) A web of disclinations observed with a single polarizer. (b) Progressive realignment of the director towards the vorticity direction at high $\dot{\gamma}$. The slow axis $\boldsymbol{\lambda}_g$ of the FWP is perpendicular to the shear direction.

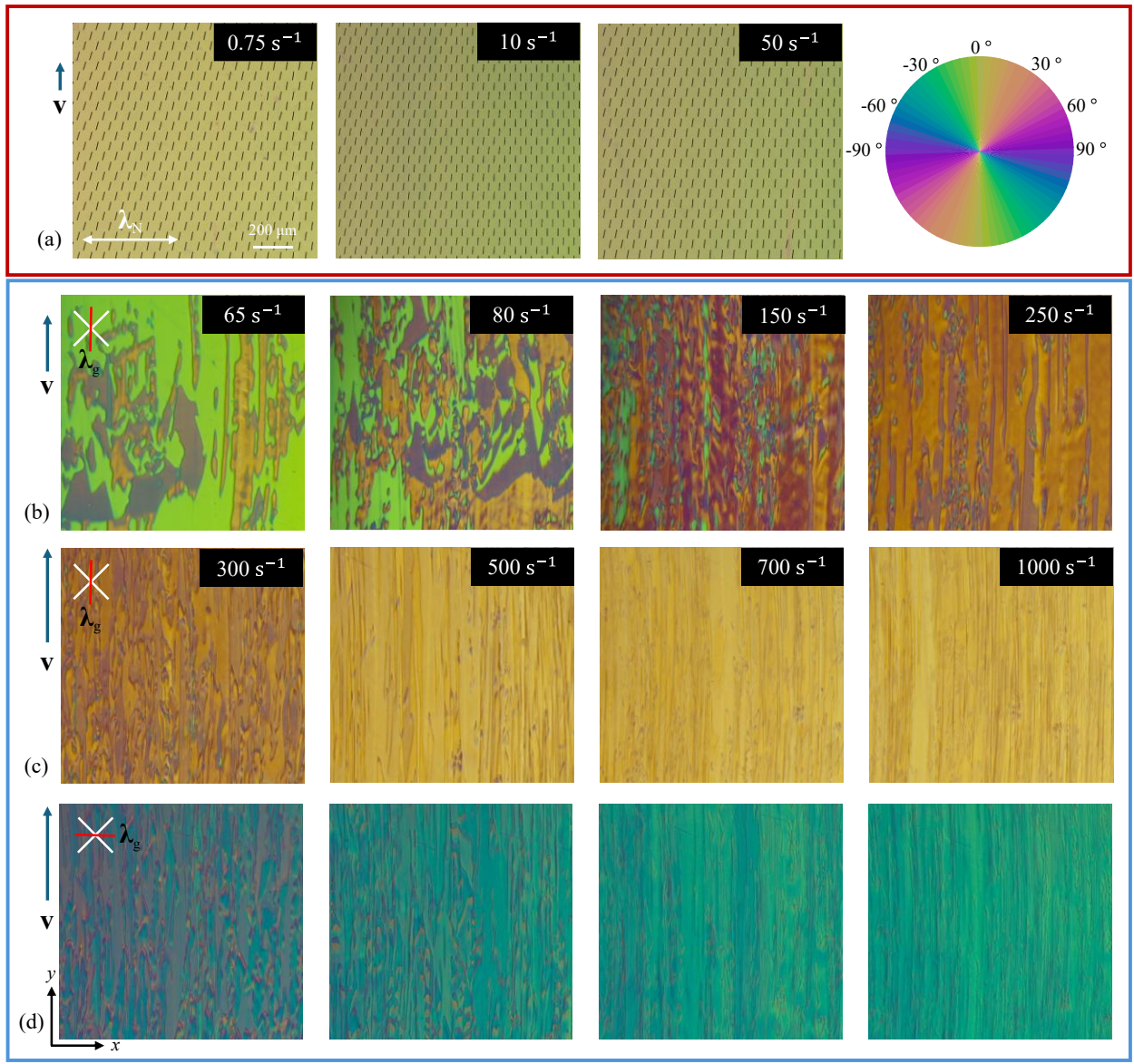

**Supplementary Fig. S3**: Structural response of the N phase of FNLC919 to shear at $T = 65\,°C$ in a 10 µm thick cell revealed in the PPM (a) and POM (b,c,d) modes of observation; the shear rate $\dot{\gamma}$ is indicated on the textures. (a) Flow-alignment at low $\dot{\gamma}$ as revealed by PPM with the N compensating cell of a thickness 6.8 µm. The color wheel describes the orientation of the optic axis (director) and the ticks in textures show the local orientation of the director. (b) Polydomain textures at the intermediate $\dot{\gamma}$ as revealed by POM with crossed polarizers and a FWP compensator. The slow axis $\lambda_g$ of the FWP shown by a red line is parallel to the shear direction. (c,d) Progressive realignment of the director towards the vorticity direction at high $\dot{\gamma}$. The slow axis $\lambda_g$ of the FWP is parallel to the shear direction in (c) and perpendicular to it in (d).

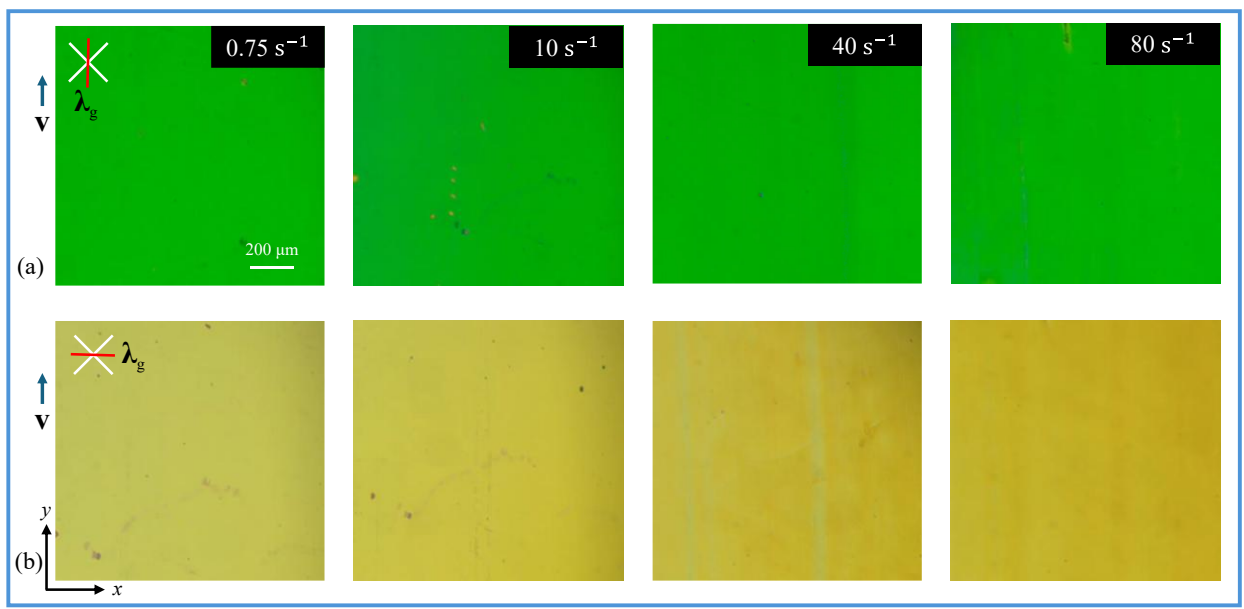

**Supplementary Fig. S4**: Structural change of DIO with an increasing shear rate in the N phase, as revealed by POM with crossed polarizers and FWP's slow axis $\lambda_g$ aligned (a) parallel to shear direction, and (b) perpendicular to it; a 10 μm thick sample and $T = 110$ °C. Shear realigns the director toward the shear plane.

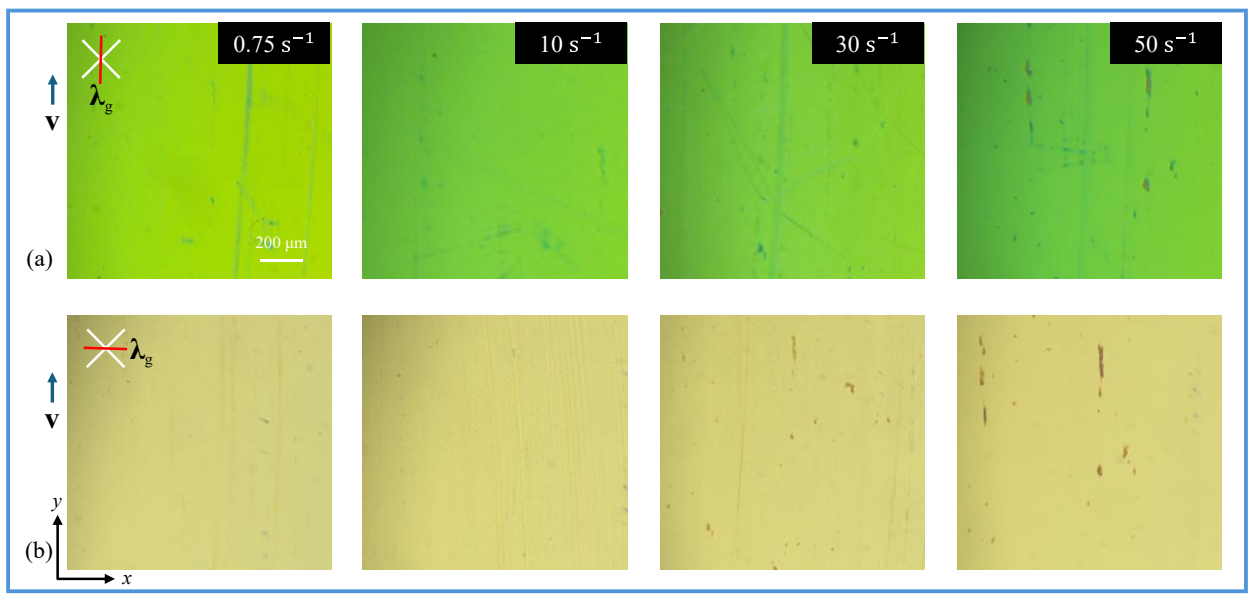

**Supplementary Fig. S5**: The realignment of the director orientation of FNLC919 along the flow in the N phase with an increasing shear rate, as revealed by POM with crossed polarizers and FWP's slow axis $\lambda_g$ aligned (a) parallel to shear direction, and (b) perpendicular to it; a 10 μm thick sample and $T = 65$ °C.

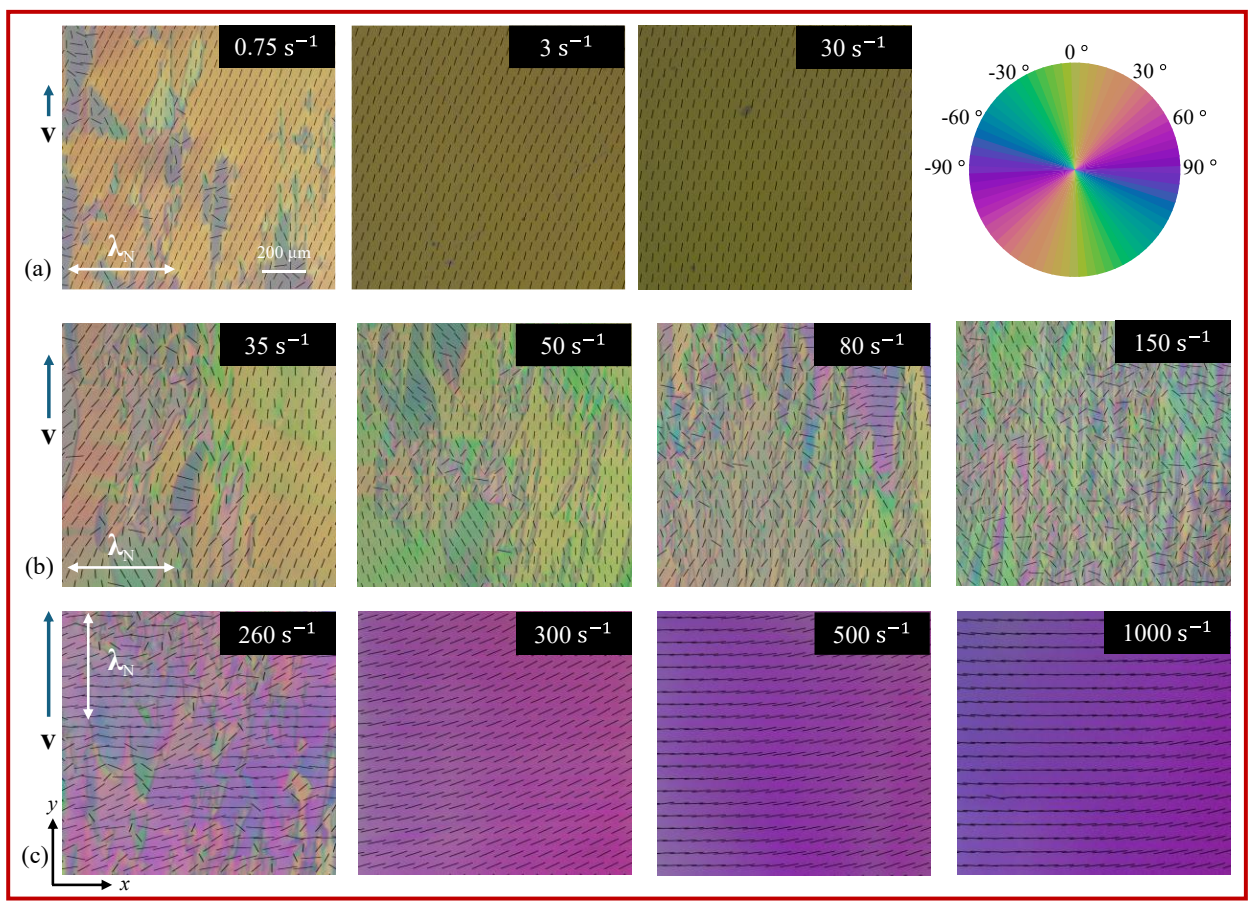

**Supplementary Fig. S6**: Structural response of the $N_F$ phase of FNLC919 to shear at $T = 25$ °C in a 10 µm thick cell revealed in the PPM modes of observation; the shear rate $\dot{\gamma}$ is indicated on the textures. (a) Flow-alignment at low $\dot{\gamma}$ as revealed by PPM with the N compensating cell of a thickness 9.6 µm. The color wheel describes the orientation of the optic axis (director) and the ticks in textures show the local orientation of the director. (b) Polydomain textures at the intermediate $\dot{\gamma}$. (c) Progressive realignment of the director towards the vorticity direction at high $\dot{\gamma}$.